\documentclass[11pt]{article}
\def\la{\langle}
\def\ra{\rangle}
\def\h{\hskip 1cm}

\def\lo{\longrightarrow}
\def\ra{\rangle}
\usepackage{graphicx}
\begin{document}
\begin{titlepage}

\vspace{4cm}

\begin{center}{\Large \bf Thermal entanglement of spins in the Heisenberg model at low temperatures}\\
\vspace{1cm} M. Asoudeh\footnote{email:asoudeh@mehr.sharif.edu},
\hspace{0.5cm} V. Karimipour \footnote{Corresponding author, email:vahid@sharif.edu}\\
\vspace{1cm} Department of Physics, Sharif University of Technology,\\
P.O. Box 11365-9161,\\ Tehran, Iran
\end{center}
\vskip 3cm

\begin{abstract}
We calculate the entanglement between two spins in the
ferromagnetic Heisenberg chain at low temperatures, and show that
when only the ground state and the one particle states are
populated, the entanglement profile is a gaussian with a
characteristic length depending on the temperature and the
coupling between spins. The magnetic field only affects the
amplitude of the profile and not its characteristic length.
\end{abstract}
\end{titlepage}

\section{Introduction}\label{intro}

Thermal entanglement in many body quantum systems has recently
attracted a lot of attention \cite{abv, bose, niel, zan, wang1,
wang2, ost1, ost2, osb}. This term is usually used to specify the
amount of entanglement which exists between two spins of a multi
spin system when the whole system is in a state of thermal
equilibrium. There are at least two very good reasons for the
interest in this problem. First, it is well known that many body
quantum systems may be more correlated than the corresponding
classical systems, the excess correlation stemming from a very
basic quantum property, namely quantum entanglement \cite{epr,
sch, bell}. Thanks to the progress in the past few years, we have
now good measures of entanglement (at least in certain limited
cases \cite{woo}) and it is most natural to quantify this excess
correlation or entanglement in many body systems.  Once a property
like entanglement becomes quantifiable, we can use it as a kind of
order parameter and study its behavior when the external
parameters of the system change. This will certainly shed light on
the nature of quantum phase transition (qualitative change in the
nature of the ground state)\cite{sad}. The second reason is that
one dimensional arrays of spins, are a natural candidate for
storing quantum information \cite{kane, hammer}. Entanglement is a
valuable resource for implementation of any kind of quantum
algorithm and quantum information protocol \cite{ben, chuang} and
one needs to find the amount of entanglement which exists between
spins or qubits and the ways to control it, when the array is in
thermal
equilibrium.\\

A prototype of a many body system is the isotropic Heisenberg spin
chain whose hamiltonian is as follows: \
\begin{equation}\label{ham}
    H = -J\sum_{l=0}^{N-1} \overrightarrow{\sigma}_l\cdot\overrightarrow{\sigma}_{l+1}+\mu B\sum_{l=0}^{N-1}
    \sigma^z_l.
\end{equation}
Here $\sigma^a_l$, where ($a=x,y$ or $z$) is a spin (Pauli)
operator acting only on site
$l$ and periodic boundary conditions are assumed.\\

In this context and to our knowledge, up to now the following
studies have been made. Nielsen \cite{niel} has studied the
entanglement of a two-spin system at finite temperature
interacting via the Heisenberg interaction. Wang \cite{wang1,
wang2} and Rigolin \cite{rigolin} have also studied the effect of
anisotropy on the thermal entanglement in a two spin system.
Arnesen, Bose and Vedral have numerically studied the variation of
two spin entanglement with temperature and magnetic field in a
few-spin isotropic Heisenberg chain \cite{abv} and Gunlycke et al.
\cite{bose} have studied the same problem in an Ising model in a
transverse field where they have observed a kind of quantum phase
transition at zero temperature when the entanglement suddenly
shows up even with an infinitesimal transverse magnetic field.
O'Connor and Wootters \cite{cw}, have calculated the maximum
possible entanglement between nearest neighbors in a
translationaly invariant state and have shown that the ground
state of the antiferromagnetic Heisenberg chain satisfies this
maximum under certain conditions. Osterloh et al \cite{ost1} have
shown that quantum phase transitions of a class of spin systems
can be characterized by the change in the entanglement between the
next and next-nearest neighbors spins. In particular they have
shown that near the point of quantum phase transition the nearest
and next-nearest neighbor entanglement exhibit  logarithmic
divergence and universal
behavior. A similar study has been done in \cite{osb}.\\

Of particular interest to us here is the work of Wang and Zanardi
\cite{zan} who have shown that in the isotropic Heisenberg model
in the absence of magnetic field, the nearest-neighbor
entanglement can be related directly to the free energy of the
model. Thus by knowing only the eigenvalues and not the
eigenstates, one can calculate the entanglement between nearest
neighbor spins. In particular they have shown that in an
Heisenberg ferromagnet there is no entanglement between nearest
neighbors if $B=0$. \\ This is a very rare and fortunate situation
where one can draw general conclusions about a problem whose
solution generally needs a knowledge of the whole spectrum, i.e.
the eigenvalues and eigenvectors. As they have correctly pointed
out, their argument is based on translational invariance of the
lattice and the $su(2)$ symmetry of the model. Once the $su(2)$
symmetry is broken, e.g. by applying a magnetic field, their
argument will not be valid anymore and one may find entanglement
in the Heisenberg ferromagnetic chain.\\ In this paper we show
that if we apply a magnetic field to a ferromagnetic Heisenberg
chain, then pairwise entanglement will develop between spins at
arbitrary given sites. An exact solution of this problem is
extremely difficult if not impossible, since as we will see it
requires the determination of the spin-spin correlation functions
over all the energy eigenstates . However at low temperatures when
only the ground state and the first excited states are populated
one can calculate the entanglement analytically. Moreover one can
now calculate the entanglement profile (entanglement between
arbitrary sites) and see its dependence on various control
parameters, like the
spin coupling, the magnetic field  and the temperature.\\
We will find that for finite but arbitrary number of spins, the
entanglement profile is a guassian with a characteristic length
depending only on the temperature and the coupling between spins.
On the other hand the magnetic field and the total number of spins
only affect the amplitude of the
profile.\\

\textbf{Remark} Contrary to the case where there is no magnetic
field, an exact proof on the presence or absence of entanglement
in the Heisenberg ferromagnetic chain is still missing. Exact
solutions for up to 4 and numerical evidence for up to 10 spins
\cite{abv} indicate that there is no entanglement in ferromagnetic
chains. If this is indeed the case then the entanglement
calculated in this paper should be looked upon not as a result of
approximation but as a result of truncation of the spectrum. That
is if we somehow prevent the higher excited states from mixing
with the low lying states of the Hiesenberg ferromagnet, then we
can produce entanglement between different remote sites and
suitably manage this entanglement. This problem may be of
practical relevance in solid state implementations of quantum
computers which should be kept at low temperature to reduce the
effect of noise. In such cases we need to know not only the
entanglement between nearest neighbor spins but also between
distant spins. The variation of entanglement with distance has not
been studied in the works mentioned above, due to the
complications in the ground and excited states of the
anti-ferromagnetic Heisenberg chain or the Ising model in
transverse field. The low lying states of the Heisenberg
ferromagnet is a simple situation where we can study this
variation.
\\
The structure of this paper is as follows: In section
\ref{general} we discuss the general form of the two body density
matrix in the Heisenberg chain.  In section \ref{heisenberg} we
review the basic properties of the low lying states of the
Heisenberg ferromagnet and in section \ref{thermal} we calculate
the explicit form of the two particle density matrix and its
concurrence which characterizes its degree of entanglement. We
conclude the paper with a discussion in section \ref{dis}.

\section{Some general properties of the two particle density matrix}\label{general}
The Hamiltonian of the Heisenberg spin chain (\ref{ham}) has the
following symmetries
\begin{equation}
    [H,S^z]=[H,\pi]=[H,T]=0,
\end{equation}
where
\begin{equation}\label{Js}
    S^{z}:=\frac{1}{2} \sum_{l=0}^{N-1} \sigma^z_l,
\end{equation}
is the third component of the total spin, and $T$ is the
translation operator
\begin{equation}\label{T}
    T|s_0, s_1, \cdots s_{N-1}\ra = |s_{1}, s_{2}, \cdots
    s_{N-1},s_0\ra.
\end{equation}
Here $|s_i\ra, \ \ \  s_i = \pm 1$ is the state of the $i$-th
spin expressed in terms of the eigenvectors of $\sigma^z$, i.e.
$\sigma^z_i|s_i\ra=s_i|s_i\ra$ and $\pi$ is the operator which
reflects the spin chain around a suitable site: i.e.
\begin{equation}\label{pi}
    \pi|s_0, s_1, \cdots s_{N-1}\ra = |s_{N-1}, s_{N-2}, \cdots
    s_{0}\ra.
\end{equation}
This symmetry means that enumerating the sites of the lattice in
the clockwise or anti-clockwise directions is immaterial for the
derivation of the properties of the lattice. The equilibrium state
of such a system at temperature $T$ is given by a density matrix
$\rho = \frac{e^{-\beta H}}{Z}$, where $\beta = \frac{1}{k_BT}$,
$k_B$ is the Boltzman constant and $Z=tr(e^{-\beta H})$ is the
partition function.\\
We are interested in the reduced density matrix and entanglement
of two spins at sites $m$ and $n$. This is given by
\begin{equation}\label{rho1}
 \rho_{m,n}=\frac{1}{Z} tr_{\widehat{m,n}}e^{-\beta H},
\end{equation}
where $\widehat{m,n}$ means that the trace is taken over all
factors in the tensor product space except those at sites $m$ and
$n$.\\
From (\ref{rho1}) we find various elements of the matrix
$\rho_{m,n}$ as

\begin{equation}\label{rho2}
  \la i,j|\rho_{m,n} |k,l\ra=\frac{1}{Z}
  tr\left(e^{-\beta H}(E_{ki}(m)E_{lj}(n))\right),
\end{equation}
where $i,j,k,l = 0, 1 $ and $E_{ki}(m)$ is the operator which acts
like $E_{ki}=|k\ra\la i|$ on site $m$ and like identity on all the
other sites. (Note that we use two equivalent notations for the
states, namely $|+\ra = |0\ra$ and $|-\ra = |1\ra$, the $\pm$
notation is usual for the description of spin states and the
$|0\ra, |1\ra$ notation is usual in description of qubit states
in quantum computing.) Expressing the matrices $E_{ki}$ in terms
of Pauli matrices and using the symmetry $[H, S^z]=0$, one can
rewrite $\rho_{m,n}$ in terms of correlation functions of spin
chains as:
\begin{equation}\label{correlation}
  \rho_{m,n}=\left(\begin{array}{cccc}
    u^{+} &  &  &  \\
     & w & z &  \\
     & z & w &  \\
     &  &  & u^{-} \
  \end{array}\right),
\end{equation}
where due to translational and reflection invariance, the
parameters $u^{\pm}$, $w$ and $z$ depend on $|m-n|$ and
\begin{eqnarray}\label{u+u-}
  u^{+} &:=&\frac{1}{4Z}tr((1+\sigma^z_m)(1+\sigma^z_n)e^{-\beta H})\cr
  u^{-} &:=&\frac{1}{4Z}tr((1-\sigma^z_m)(1-\sigma^z_n)e^{-\beta
  H})\cr
  w&:=& \frac{1}{4Z}tr((1-\sigma^z_m)(1+\sigma^z_n)e^{-\beta
  H})\cr
  z&:=& \frac{1}{Z} tr(\sigma_m^-\sigma_n^+ e^{-\beta H}).
\end{eqnarray}
For notational convenience we do not show the explicit dependence
on $|m-n|$ in the parameters of $\rho_{m,n}$. The equality of the
elements $\rho_{01,10}$ and $\rho_{10,01}$, (the reality of $z$)
is a consequence of the $\pi$ symmetry and translational
symmetry. This is pointed out by Wang and Zanardi in \cite{zan},
although they explicitly derive this property by the symmetry
$[(\sigma^x)^{\otimes N},H]=0$ which exists only in the absence
of magnetic field. To see this we temporarily insert the site
dependence in the parameters and note that
\begin{eqnarray}\label{realityofz}
  z(m-n)&=&tr(\rho \sigma^-_m\sigma^+_n)=tr(\pi\rho\pi
  \sigma^-_{m}\sigma^+_{n})\cr &=& tr(\rho \pi\sigma^-_m\sigma^+_n\pi)=
  tr(\rho
  \sigma^-_{N-m}\sigma^+_{N-n})\cr &=& tr(\rho
  \sigma^+_{N-n}\sigma^-_{N-m})=
  z^*(N-n-(N-m))\cr &=&z^*(m-n).
\end{eqnarray}
The same argument shows why $\rho_{01,01}=\rho_{10,10}$.\\
 The eigenvalues of
this density matrix (which is positive) are $u^{\pm}$ , and $w\pm
z$, which implies that the $u^{\pm}$ and $w$ are positive, while
the sign of $z$ is not determined.\\
The entanglement of the two spins at sites $m$ and $n$ is
determined by the concurrence $C(|m-n|)$ defined as
\begin{equation}\label{con}
  C(|m-n|) = max \{ 0, \lambda_1-\lambda_2-\lambda_3-\lambda_4\},
\end{equation}
where  $\lambda_1, \lambda_2,\lambda_3, $ and $\lambda_4 $ are the
positive square roots of the eigenvalues of the matrix
$\rho_{m,n}\tilde{\rho}_{m,n}$ in decreasing order. The matrix
$\tilde{\rho}$ is defined as
\begin{equation}\label{rho6} \tilde{\rho}_{m,n} = (\sigma^y \otimes
\sigma^y)\rho_{m,n}^*(\sigma^y \otimes \sigma^y),
\end{equation}
where $*$ denotes complex conjugation in the computational basis.
It turns out that \cite{cw} $C(|m-n|)$ is given by
\begin{equation}\label{Cmn}
  C(|m-n|) = 2 \ {\rm {max}}\ (0, |z|-\sqrt{u^+u^-}).
\end{equation}
\section{The low lying states of the Heisenberg ferromagnet}\label{heisenberg}
The Hamiltonian (\ref{ham})
 can be rewritten as
\begin{equation}\label{ham3}
    H = JN - 2J \sum_{l=0}^{N-1} P_{l,l+1}+\mu B \sum_{l=0}^{N-1}
    \sigma^z_l,
\end{equation}
where $P_{l,l+1}$ is the permutation operator acting on sites $l$
and $l+1$, $P|s,s'\ra = |s',s\ra$.\\
For the ferromagnetic chain ($J > 0$), the ground state of $H$ is
the disentangled state
\begin{equation}\label{gnd}
   |\epsilon_0\ra = |-, -, -, \cdots , -, -, -\ra,
\end{equation}
where all the spins are down, with energy $ \epsilon_0 = -(J+\mu
B)N$.\\ \\ The first excited states are linear combination of one
particle states
\begin{equation}\label{1par}
    |\psi\ra = \sum_{k}c_k |k\ra,
    \end{equation}
where
\begin{equation}\label{k1}
    |k\ra := |-,-,-,\cdots ,+,\cdots ,-,-,-\ra,
\end{equation}
in which only the spin in the $k-$th site is up. \\
From
(\ref{ham3}) we find
\begin{equation}\label{k2}
    H|k\ra = -2J|k-1\ra - 2J|k+1\ra +
    \left(J(4-N)+\mu B(2-N)\right)|k\ra.
\end{equation}
leading to the recursion relations for the coefficients
\begin{equation}\label{rec}
    -2J(c_{k-1}+c_{k+1})+ \left(J(4-N)+\mu B(2-N)\right)c_k = \lambda
    c_k
\end{equation}
whose solution together with the periodic boundary condition gives
us the final form of the first excited states labeled by
$|\psi_s\ra $ with $s = 0 \cdots N-1$:
\begin{equation}\label{psi}
    |\psi_s\ra = \frac{1}{\sqrt{N}}\sum_{0}^{N-1} e^{\frac{2\pi i
    sk}{N}}|k\ra,
\end{equation}
with energies
\begin{equation}\label{lam}
\lambda_s =\epsilon_0+ 2\mu B + 8J \sin^2 \frac{\pi s}{N} .
\end{equation}
The lowest energy in this band belongs to the flat wave ($s=0$)
which is higher than the ground state energy by $2\mu B$. In the
classical picture this is the energy difference due to the
flipping of one spin in the magnetic field and there is no
contribution from the interaction of spins. One can guess that
the lowest energy in the two particle sector also corresponds to a
flat wave, namely to a state
\begin{equation}\label{flat2}
  |\Psi\ra = \frac{1}{\sqrt{\left(\begin{array}{c}
    N \\
    2 \
  \end{array}\right)}}\sum_{0\leq i< j\leq N-1} |i,j\ra,
\end{equation}
where $|i,j\ra$ is a state in which only the spins at sites $i$
and $j$ are up. It is easy to see that this is indeed an
eigenstate of
the Hamiltonian with an energy $\epsilon_0 + 4\mu B$. \\
Thus if $\frac{2\mu B}{kT}>> 1$, one may assume that only the
ground state and the first excited states will be populated. We
will assume that the magnetic field is strong enough and the
temperature is low enough so that this condition is fulfilled.

\section{Thermal entanglement of two spins}\label{thermal} We will
calculate the density matrix (\ref{rho1}) at low temperatures by
taking into account only the ground state and the one-particle
eigenstates. At low temperature we approximate various quantities
in (\ref{u+u-}) as follows:
\begin{equation}\label{approx}
  tr(Ae^{-\beta H}) \approx \la\epsilon_0 |A|\epsilon_0\ra e^{-\beta \epsilon_0} +
   \sum_{s=0}^{N-1}\la \psi_s|A|\psi_s\ra e^{-\beta \lambda_s}.
\end{equation}

The operator $A$ takes one of the following forms,
$(1\pm\sigma^z_m)(1\pm\sigma^z_n)$,
$(1+\sigma^z_m)(1-\sigma^z_n)$, or $\sigma^-_m \sigma^+_n$. \\
We need the relevant matrix elements over the one particle
states. It is easily verified that

\begin{eqnarray}\label{1}
\frac{1}{2}(I+\sigma_n^z)|k\ra &=& \delta_{n, k}|k\ra,\cr
\frac{1}{2}(I-\sigma_n^z)|k\ra &=& (1-\delta_{n,k})|k\ra,\cr
\sigma_n^-|k\ra &=& \delta_{n,k}|k\ra,
\end{eqnarray}
from which we obtain the following matrix elements, where we have
used the fact that $m\ne n$:
\begin{eqnarray}\label{z+z+psi}
\frac{1}{4}\la \psi_s |(1+\sigma^z_m)(1+\sigma^z_n)|\psi_s\ra &=&
0,\cr\frac{1}{4} \la \psi_s
|(1+\sigma^z_m)(1-\sigma^z_n)|\psi_s\ra &=& \frac{1}{N}\sum_{k,
l=0}^{N-1} e^{\frac{2\pi i(l-k)s}{N}} \delta_{m, k}\delta_{k, l}
= \frac{1}{N},\cr  \frac{1}{4}\la \psi_s
|(1-\sigma^z_m)(1-\sigma^z_n)|\psi_s\ra &=& 1-\frac{2}{N}\cr \la
\psi_s |\sigma^-_m\sigma^+_n|\psi_s\ra  &=& \frac{1}{N}
e^{\frac{+2\pi i(m-n)s}{N}}.
\end{eqnarray}
From  (\ref{u+u-}) and (\ref{z+z+psi}) we find values of different
matrix elements of $\rho_{m,n}$ as follows:

\begin{eqnarray}\label{abd}
u^+ &=& 0 \cr w &=& \frac{1}{Z} \frac{\sum_{s=0}^{N-1} e^{-\beta
    \lambda_s}}{N}\cr
 u^- &=& \frac{1}{Z} \left(e^{-\beta\lambda_0} +
(1-\frac{2}{N})\sum_{s=0}^{N-1}e^{-\beta \lambda_s} \right)\cr z
&=& \frac{1}{N} \frac{1}{Z} \sum_{s=0}^{N-1} e^{\frac{2\pi
i(m-n)s}{N}-\beta \lambda_s}.
\end{eqnarray}

The partition function at this level of approximation is given by

\begin{equation}\label{par}
    Z \approx e^{-\beta \epsilon_0} + \sum_{k=0}^{N-1}e^{-\beta
    \lambda_s} = e^{-\beta\epsilon_0} + e^{-\beta(\epsilon_0+2\mu
    B)}\sum_{s=0}^{N-1} e^{-8\beta J \sin^2 \frac{\pi s}{N}}.
\end{equation}

For large $N$ we can approximate the sums in the above formulas by
integrals and at low temperatures (large $\beta$) we can evaluate
the integrals by saddle point approximation. Thus

\begin{eqnarray}\label{sum1}
    \sum_{k=0}^{N-1}e^{-\beta
    \lambda_s} &=& e^{-\beta(\epsilon_0+2\mu
    B)}\sum_{s=0}^{N-1} e^{-8\beta J \sin^2 \frac{\pi s}{N}}\cr
&\approx& e^{-\beta(\epsilon_0+2\mu B)}
\int_{\frac{-N}{2}}^{\frac{N}{2}}
      e^{-8\beta J \sin^2 \frac{\pi s}{N}} ds \cr
      &\approx&  e^{-\beta(\epsilon_0+2\mu B)} N \sqrt{\frac{1}{8\pi\beta J}}
      .
\end{eqnarray}
The partition function will be
\begin{equation}\label{partition2}
  Z \approx e^{-\beta\epsilon_0} \left(1+e^{-2\beta\mu B}N\frac{1}{\sqrt{8\pi\beta
  J}}\right).
\end{equation}
 We also find
\begin{eqnarray}\label{csad}
     \sum_{s=0}^{N-1} e^{-\frac{2\pi i(n-m)s}{N} - \beta
     \lambda_s}&\approx& e^{-\beta(\epsilon_0+2\mu B)}
    \int_{\frac{-N}{2}}^{\frac{N}{2}} e^{-8\beta J \sin^2 \frac{\pi s}{N}-\frac{-2\pi i(n-m)s}{N}} ds \cr
      &\approx& e^{-\beta(\epsilon_0+2\mu B)} N \sqrt{\frac{1}{8\pi\beta
      J}}e^{-\frac{(n-m)^2}{8\beta J}}.
\end{eqnarray}
Putting all this together we find
\begin{equation}\label{rhomnfinal}
  \rho_{m,n} = \frac{1}{N + \sqrt{8\pi \beta J} e^{2\beta \mu
  B}}\left(\begin{array}{cccc}
    0 &  &  &  \\
     & 1 & e^{-\frac{(n-m)^2}{8\beta J}} &  \\
     & e^{-\frac{(n-m)^2}{8\beta J}} & 1 &  \\
     &  &  & N-2 + \sqrt{8\pi \beta J} e^{2\beta \mu B}
  \end{array}\right).
\end{equation}

\begin{figure}
\centering
    \includegraphics[width=12cm,height=14cm,angle=-90]{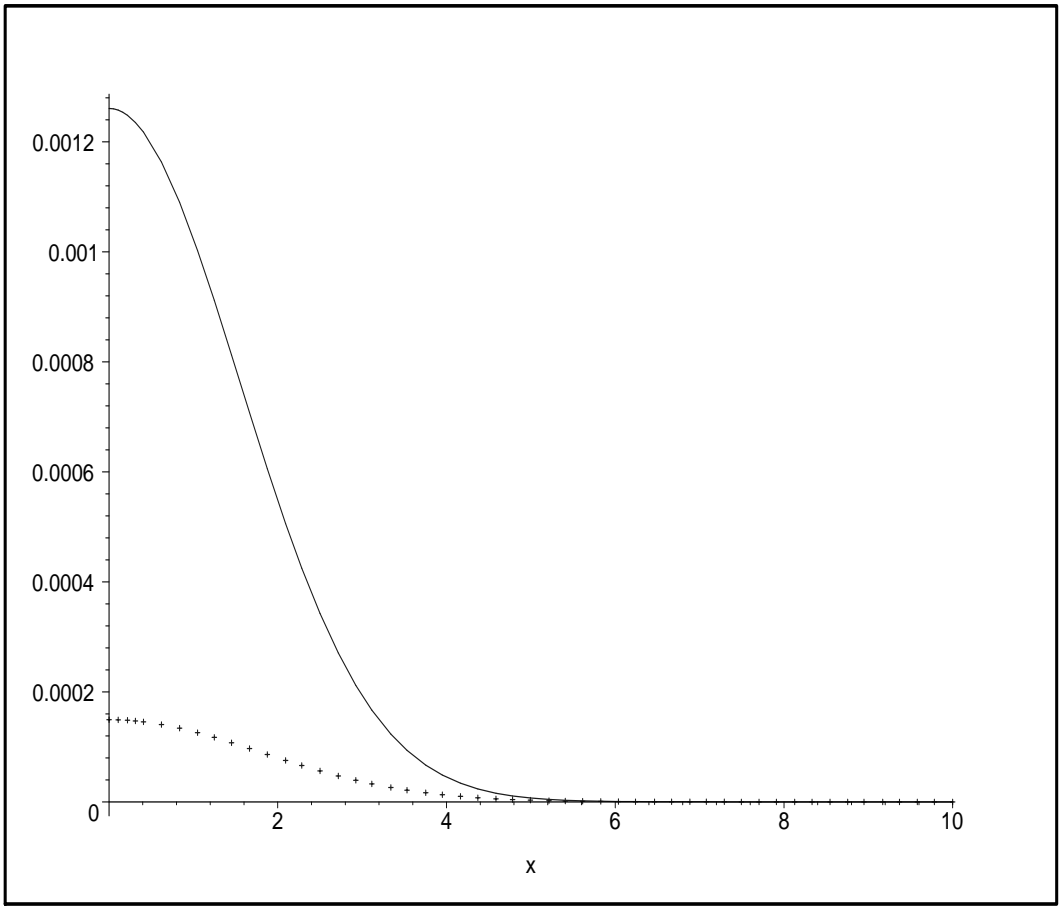}
    \caption{The concurrence $C$ as a function of $x=|m-n|$, the distance between spins,
     for a spin chain of $N=20$ spins at two different temperatures.
    Solid line ($\frac{\mu B}{kT}=3,\frac{J}{kT}=0.6 $), dashed line ($\frac{\mu B}{kT}=4,\frac{J}{kT}=0.8 $)}.\label{profile}
\end{figure}

Now that we have the two particle density matrices, we can
calculate the entanglement of two spins at an arbitrary distance.
From (\ref{Cmn}) we find
\begin{equation}
  C(|m-n|)=2|z|=\frac{2}{N + \sqrt{8\pi \beta J} e^{2\beta \mu
  B}} e^{-\frac{(n-m)^2}{8\beta J}},
\end{equation}
which can be rewritten as
\begin{equation}\label{Csimple}
  C(|m-n|)=C_0 e^{-\frac{1}{2}\frac{(m-n)^2}{l^2}},
\end{equation}
where
\begin{equation}\label{Cmax}
  C_0:=\frac{2}{N + \sqrt{8\pi \beta J}e^{2\beta \mu B}} \h {\rm{and}}\h
  l = 2 \sqrt{\frac{J}{kT}}
\end{equation}
denote respectively a scale of the value of entanglement and a
kind of entanglement length which determines a scale of length
over which the distant spins are still entangled. It is seen that
the magnetic field has no effect on the entanglement length and
only controls the entanglement amplitude. The higher the magnetic
field the lower the value of $C_{0}$. This is understandable
since a high value of the magnetic field orders all the spin more
in the direction of the $z$ axis in favor of the disentangled
ground state. On the other hand it is the coupling between
adjacent spins that entangles distant spins, hence the
entanglement length is only affected by $J$ and $T$. Raising the
temperature has a destructive effect on the entanglement length
since thermal fluctuations tend to move the system away from the
quantum regime. Finally we note that in the thermodynamic limit
when $N\lo \infty$, the concurrence $C_{0}$ tends to zero, so
that even the nearest neighbor sites will not be entangled
anymore. Figure \ref{profile} shows the entanglement profile for
a spin chain of $N=20$ spins at two different temperatures.

\section{Discussion}\label{dis}
As Wang and Zanardi \cite{zan} have shown, in the Heisenberg
ferromagnet in the absence of a magnetic field there can be no
thermal entanglement between nearest neighbors spins.  Their
argument is based on the $su(2)$ symmetry of the Hamiltonian. In
the presence of a magnetic field this symmetry breaks down to a
$u(1)$ symmetry and one can only derive entanglement by explicitly
calculating the relevant thermal correlation functions of the
model. We have shown that at low temperatures or high magnetic
fields where to a good approximation only the ground state and
the one particle states are populated, entanglement can exist
between different sites. We have shown that the entanglement
profile is a gaussian with an amplitude depending on the magnetic
field and temperature and a characteristic length
depending only on temperature and the coupling between spins. \\
Another important point is that in the previous works the pairwise
entanglement has been studied in more complicated situations and
for making connections to quantum phase transitions, but only for
nearest neighbors. By considering the forromagnetic case and by
restricting to the low lying states, we have been able to see how
entanglement varies with distance. Knowing this variation may be
important for the implementation of quantum protocols on arrays
of qubits in any candidate for implementation of quantum
computers. An analogous study even for the ground state of the
anti-ferromagnetic spin chain is extremely difficult, since it
requires the calculation of correlation functions for arbitrary
distances and up to now the exact ground state correlation
functions of the anti-ferromagnetic model have been calculated
for distances of only up to 4 sites \cite{tak}.

\section{Acknowledgement} We thank Andreas Osterloh for bringing references \cite{ost1,ost2,osb,jin} to our attention.

\end{document}